\documentclass[preprint]{aastex}
\usepackage{emulateapj5}
\def\stacksymbols #1#2#3#4{\def\theguybelow{#2}
        \def\verticalposition{\lower#3pt}
        \def\spacingwithinsymbol{\baselineskip0pt\lineskip#4pt}
        \mathrel{\mathpalette\intermediary#1}}
\def\intermediary #1#2{\verticalposition\vbox{\spacingwithinsymbol
        \everycr={}\tabskip0pt
        \halign{$\mathsurround0pt#1\hfil##\hfil$\crcr#2\crcr
                \theguybelow\crcr}}}
\def\lta{\stacksymbols{<}{\sim}{2.5}{.2}}
\def\gta{\stacksymbols{>}{\sim}{3}{.5}}

\begin{document}

\title{TIME-DEPENDENT CIRCULATION FLOWS: IRON ENRICHMENT 
IN COOLING FLOWS WITH HEATED RETURN FLOWS}


\author{William G. Mathews\altaffilmark{1}, 
Fabrizio Brighenti\altaffilmark{1}\altaffilmark{2},
\& David A. Buote\altaffilmark{3}}

\altaffiltext{1}{University of California Observatories/Lick Observatory,
Department of Astronomy and Astrophysics,
University of California, Santa Cruz, CA 95064, 
mathews@ucolick.org}

\altaffiltext{2}{Dipartimento di Astronomia,
Universit\`a di Bologna,
via Ranzani 1,
Bologna 40127, Italy, 
brighenti@bo.astro.it}

\altaffiltext{3}{University of California at Irvine, 
Dept. of Physics \& Astronomy, 4129 Frederick Reimes Hall,
Irvine, CA 92697, buote@uci.edu}






\vskip .2in

\begin{abstract}
We describe a new type of dynamical model for hot gas 
in galaxy groups and clusters in which gas moves 
simultaneously in both radial directions.
The observational motivations for this type of flow are 
compelling. 
X-ray spectra indicate that little or no gas is cooling 
to low temperatures.
Bubbles of hot gas typically appear in Chandra X-ray images 
and XMM X-ray spectra within $\sim 50$ kpc of the central 
elliptical galaxy.
These bubbles must be buoyant. 
Furthermore, the elemental composition and 
total mass of gas-phase iron observed within $\sim 100$
kpc of the center can be understood as the accumulated outflow
of most or all of the iron produced by Type Ia supernovae
in the central galaxy over time. 
This gaseous iron has been circulating for many Gyrs, 
unable to cool.
As dense inflowing gas cools, it produces a 
positive central temperature gradient, 
a characteristic feature of normal cooling flows.
This gas dominates the local X-ray spectrum 
but shares the total available volume with 
centrally heated, outflowing gas.
Circulating flows eventually cool catastrophically if 
the outward flowing gas transports mass but no heat; 
to maintain the circulation both mass and energy must be supplied 
to the inflowing gas over a large volume,
extending to the cooling radius.
The rapid radial recirculation of gas within $\sim50$ kpc 
results in a flat core in the gas iron abundance, 
similar to many group and cluster observations.
We believe the circulation flows described
here are the first gasdynamic, long-term evolutionary
models that are in good agreement
with all essential features observed in the hot gas:
little or no gas cools as required by
XMM spectra, the gas temperature increases outward near the center,
and the gaseous iron abundance is about solar near the center
and decreases outward.
\end{abstract}



\section{Introduction}

Now that we have become accustomed to referring to the 
hot gas in groups and clusters of galaxies as ``cooling flows'', 
no cooling gas has been observed. 
Although energy is lost by X-ray emission, 
the absence of spectral evidence for gas cooling to low temperatures 
has been well-documented in many observational papers
(e.g. Peterson et al. 2001; Xu et al. 2002). 
The hot gas in groups and clusters of galaxies may cool, 
but at a much lower rate than previously thought.
This has led to numerous theoretical models in which the gas is 
assumed to be 
reheated in some fashion -- by powerful radio jets, by rising 
bubbles, by shocks, by magic etc. 
Most or all of these models have been unsatisfactory at some level.
The most common difficulty of heated flows is 
not reproducing the observed radial temperature and density 
profiles (e.g. Brighenti \& Mathews 2002, 2003),
which are smooth and approximately similar 
for galaxy clusters of all scales. 
Less theoretical effort has been expended in understanding 
the radial abundance gradients in the hot gas, 
but it is fair to say that little progress has been made so far 
on this important problem.

In this paper we show that a combination of heating and 
radial mass redistribution can satisfy these stringent 
observational constraints and simultaneously explain the 
observed radial metallicity profiles. 
While the flows described here 
retain some useful features of traditional 
cooling flows, we refer to them as ``circulation flows'' since 
gas flows in both directions simultaneously. 
Cooling to low temperatures is greatly or entirely eliminated.
The primary source of heat that drives the circulation is 
assumed to be moderate AGN activity in the 
cores of luminous elliptical galaxies that 
reside at or near the center of the diffuse X-ray emission 
in groups and poor clusters.

Our theoretical and observational studies of the 
viralized hot gas in galaxy clusters 
have been motivated in several ways.
First, we have maintained that hot gas 
in galaxy groups is more likely 
to reveal the physical nature and dynamical evolution 
of the gas than hotter gas in rich clusters.
We assume that 
some of the groups are old and, if cosmically isolated, 
are less likely to have been upset and mixed by recent mergers 
or powerful radio sources. 
Second, we have argued that 
the radial variation of the gaseous iron abundance in these
groups retains an imprint of the origin and 
dynamical evolution of the gas since it received its iron.
Finally, at the relatively low temperatures of gas in 
galaxy groups, $kT \sim 1$ keV, the influence 
of thermal conductivity $\kappa \propto T^{5/2}$ 
on the scale of the flow can be ignored.

In many galaxy groups and clusters the gas-phase iron abundance 
has a noticeable peak centered on the central elliptical 
galaxy that extends out to $\sim 100$ kpc.
The iron abundance increases to $\sim$solar at the center.
XMM spectra indicate that 70 - 80 percent of this iron 
was formed in Type Ia supernovae. 
Of particular interest for our discussion here 
is the realization that the 
total amount of gas-phase iron in this central region is
comparable to the total amount of iron that could be 
created by all Type Ia supernovae in the central
galaxy since its stars were formed. 
This is only possible if little or none of the iron-enriched 
gas has cooled.
It is also significant that the region of enhanced iron 
emission around the central galaxy is much larger in 
extent than the half-light radius of the optical galaxy.

To explain these important features, 
we describe a simple time-dependent model for the 
long term evolution of hot gas in galaxy groups in 
which gas enriched by Type Ia supernovae inside the central 
galaxy is heated and buoyantly transported far 
into the surrounding gas where it is stored over time. 
Meanwhile, most of the hot X-ray emitting 
gas loses energy and flows inward, as in a standard 
cooling flow.
The cooling inflow receives both iron-enriched gas and energy
from clouds or bubbles of heated gas that are moving outward.
To avoid flows with catastrophic cooling
or discordant temperature profiles, it is essential that the
inflowing gas receive both energy and mass from the
central regions.
The physical nature of this heating -- $PdV$ compression,
sound wave dissipation, weak shocks, etc, -- is left 
unspecified for the time being. 
For successful flows the mass and energy deposition must 
be spatially broad, not concentrated near a single radius 
in the flow. 
This can be accomplished if inflowing 
gas arriving near the central galaxy is heated to a 
variety of entropy levels and ultimately floats upward 
approximately to that radius where its entropy matches 
that of the surrounding, inflowing gas. 

The discussion below is a time-dependent generalization of the 
steady state circulation flows that we discussed in 
a previous paper (Mathews et al. 2003). 
Flows in which gas moves in both radial 
directions at each radius 
are notoriously difficult to reproduce faithfully 
even with the most sophisticated multidimensional numerical codes 
because numerical diffusion across the 
Eulerian grid blurs the distinction 
between the counter-streaming flows.
For this reason, in this initial study of the 
time-dependent evolution 
of radially circulating gas, we study the evolution of the 
inward flowing gas with a standard Eulerian code,
but simplify our treatment of the outflowing gas 
that is difficult to resolve on computational grids.
The physical processes by which gas near the 
central elliptical galaxy is heated are left unspecified.
Some of the less important aspects of bubble dynamics 
that we discussed in our 
study of steady state circulation flows -- 
such as momentum exchange by the drag interaction
of rising bubbles, expansion 
of individual bubbles, etc -- are either not treated in detail 
here or are represented 
by source terms in the equations for the inflowing gas. 
Since the dynamical time for buoyant gas 
is much less than the radiative cooling 
time, we assume that heated outflowing clouds 
move rapidly to their final destination.
Nevertheless, 
we do not expect that the detailed inclusion of these 
complications will alter the basic character -- or the 
success -- of the circulation flows described here.

We show that time-dependent circulation flows 
can quite naturally produce all of the major observed radial 
profiles in the hot gas: its temperature, density and 
metallicity. 
In addition to these important attributes, our circulation 
flows are compatible with the X-ray observation 
that very little if any gas cools to low temperatures.

\section{Observations of Iron-enriched Hot Gas in Galaxy Groups}

A common feature of well-studied X-ray luminous galaxy groups is 
an extended region of enhanced iron emission 
surrounding the central elliptical
(e.g. Buote et al. 2003a,b; 
Sun et al. 2003; 
Xue, B\"ohringer \& Matsushita 2004; 
Kim \& Fabbiano 2004).
Typically, the iron abundance is near-solar within 30 - 40 kpc
of the center and decreases outward within 100 - 150 kpc.
The ratio of $\alpha$/Fe elements suggest that a large fraction 
of this iron has been produced by Type Ia supernovae (SNIae).
Today the stars in this region 
from which the SNIae presumably occurred are located 
almost entirely within 
the central E galaxy with half-light radius $R_e \sim 10$ kpc,
much smaller than the scale of iron enrichment in the 
surrounding gas.

For example, the X-ray luminous group NGC 5044 consists of 
a large elliptical galaxy surrounded by 
a cloud of much smaller galaxies extending 
to at least 400 - 500 kpc.
At a distance of 33 Mpc, the central galaxy has 
luminosity $L_B = 4.5 \times 10^{10}$ 
$L_{B,\odot}$, effective radius $R_e = 10.0$ kpc, and total stellar 
mass $M_{*t} = \Upsilon_B L_B = 3.4 \times 10^{11}$ $M_{\odot}$
where $\Upsilon_B = 7.5$ is the stellar mass to light ratio.

The galaxy is surrounded by a huge cloud of hot gas extending
to at least 300 kpc.
If the hot gas is assumed to have a single phase, 
its temperature rises from $kT = 0.7$ keV at the center 
to a maximum of $kT \approx 1.3$ keV at 40-80 kpc, 
then decreases slowly beyond 
(Buote et al. 2003a).
Similar temperature profiles are observed in many other groups
and poor clusters 
(Sun et al. 2003; Loken et al. 2002).
However, within about 30 kpc the X-ray spectrum is fit better 
with two or more gas temperatures
spanning the range $0.7 \lta kT \lta 1.3$ keV
(Buote et al. 2003a).
The low temperature component at $kT \sim 0.7$ keV dominates 
the emission in this region but only occupies 
$\sim 0.5$ of the volume out to 20 or 30 kpc.
This suggests that the remaining volume is occupied by the 
hotter gas which must be buoyant if it is in 
pressure equilibrium. 
This conclusion is directly supported by the 
pronounced spatial irregularities visible in the Chandra 
image of NGC 5044 (Buote et al. 2003a).

The iron abundance in NGC 5044 decreases from $\sim 1.3$ solar 
near the center to $\sim 0.1$ solar at r = 150 kpc and remains 
approximately uniform beyond 
(Buote et al. 2003b; Buote, Brighenti \& Mathews 2004).
The total mass of iron observed within 100 kpc is 
$1.2 \times 10^8$ $M_{\odot}$ and $\alpha$/Fe abundance 
ratios indicate that 70 to 80 percent of this iron 
has been produced in Type Ia supernovae.

It is interesting to estimate the total mass of iron 
produced by SNIae in the central E galaxy over cosmic time,
\begin{equation}
M_{Fe} = \int_{t_o}^{t_n} SNu(t) {1 \over 100}
{L_B \over 10^{10} L_{B,\odot}} y_{Ia,Fe} 10^9 dt
\end{equation}
in $M_{\odot}$ 
(Mathews 1989) 
where $t$ is in Gyrs and $t_n = 13.7$ Gyrs is the current
age of the universe. 
The iron yield of each Type Ia supernova is $y_{Ia,Fe} = 0.7$ 
$M_{\odot}$.
We assume a Type Ia supernova rate of 
\begin{equation}
SNu(t) = 0.16 (t_n /t).
\end{equation}
in SNu units (rate per 100 years per $10^{10} L_{B,\odot}$).
The current Type Ia rate in early type galaxies is 
$0.16 \pm 0.06$ (Cappellaro, Evans \& Turatto 1999).
The $SNu \propto t^{-1}$ variation is consistent 
with the $SNu \propto t^{-1.2}$ variation observed by 
Pain et al. (2002) out to redshift 0.5, although the 
exponent 1.2 is uncertain.
Suppose for example that most of the stars in NGC 5044 formed 
at cosmic time $t_f = 1.65$ Gyrs (redshift $z = 4$)
and a 1 Gyr delay preceded the first 
Type Ia supernova (Gal-Yam \& Maoz 2004).
In this case
the integration above begins at time $t_o = 2.65$ Gyrs
corresponding to redshift $z \sim 2.5$. 
The integration 
predicts that $1.1 \times 10^8$ $M_{\odot}$ of iron 
has been created by all SNIae in the central elliptical NGC 5044.

It is remarkable that 
this estimate is comparable to the total mass of iron 
observed in the central iron peak out to 100 kpc, 
$M_{Fe,obs} \approx 1.2 \times 10^8$ $M_{\odot}$, 
suggesting that rather little SNIa iron is currently 
in the stars. 
At $r \gta 100$ kpc the iron abundance in NGC 5044 
plateaus at $z_{Fe} \sim 0.1$ solar
(Buote, Brighenti \& Mathews 2004) 
so it is natural to
assume that the background gas was pre-enriched everywhere  
by Type II supernovae in the early universe 
to $z_{Fe} \approx 0.1$ solar.
The total iron mass from this background within 
100 kpc is only $0.2 \times 10^8$ $M_{\odot}$,
suggesting that 80 percent of the iron in the central 
peak is from Type Ia supernovae, which is consistent
with X-ray spectra. 
While the uncertainty of past supernova rates certainly tempers
our estimate of the total iron mass from SNIae,
its similarity to the observed mass 
has important implications for the global gas dynamics. 
Very recently, De Grandi et al. (2004) have shown that the 
mass of the excess iron surrounding cluster-centered 
elliptical galaxies 
correlates with their optical luminosity, just as we would expect 
from the integration above. 
Furthermore, most of the gas-phase iron in the central
iron peaks of clusters and groups
is observed to have an SNIa abundance signature
within about 100 kpc of the central elliptical galaxy
(De Grandi \& Molendi 2001). 

One possible explanation for the extended 
regions of iron enrichment is that the iron was created in SNIae 
within the central E galaxy, $r \sim R_e = 10$ kpc, 
but was subsequently distributed outward by merging events 
into the much larger region where 
it is observed today.
Such mixing due to gas-phase mergers 
would need to be accompanied by an 
increased entropy if the iron-rich gas is to 
remain at a larger radius.
If a merger radially mixes the gas without strong heating, 
after a dynamical time the gas will return
to its original radius in the entropy structure 
and the original metallicity gradient will be restored.
However, we expect that most of the shock energy dissipated 
(and entropy created) 
in mergers occurs closer to the virial radius which 
is very much larger than the $r \lta 100$ kpc region 
enriched in iron. 
Central increases in gas-phase metallicity 
are only observed in the most relaxed groups and clusters -- 
they are clearly an indication of an old system. 
If mixing due to mergers occurred long ago, 
not enough SNIa iron would have been available to 
account for current observations. 

In our previous gasdynamical studies including iron 
enrichment (reviewed in Mathews \& Brighenti 2003), 
we found that the radial distribution of 
gas-phase iron from both Type II and 
early-time Type Ia supernovae can extend well beyond the 
optical radius $R_e$ of the central elliptical galaxy.
This spatial spreading of the iron can result from early 
galactic winds if the central elliptical formed monolithically,
or from SNIae in stars in the central elliptical 
located well beyond $R_e$ where 
both the stellar and gas densities are low.
Nevertheless, these models cannot spread the iron to 
radii as large as $\sim 100$ kpc where it is typically observed.

In a recent paper (Buote, Brighenti \& Mathews 2004) 
we have shown that very little of the gas-phase iron in 
the NGC 5044 group has been supplied by outflows from 
the large number of low luminosity group member galaxies.
The radial distribution of stellar mass in 
these non-central group galaxies
is approximately constant within the $\sim 100$ kpc 
region where the gas iron enrichment has a strong gradient.
It is clear from our arguments in that paper that the 
mass of iron in the 
central peak within $\sim100$ kpc 
cannot be ascribed to starburst outflows 
or winds associated with other members of the NGC 5044 group.

It is difficult to escape the conclusion that a very large 
fraction of the iron in $r \lta 100$ kpc was produced 
by Type Ia supernovae associated with stars that are
currently within the large central elliptical,
but where were these stars when the iron was produced? 
Perhaps the extended iron enrichment occurred 
as the stars within the central E galaxy were assembled.
This seems unlikely, however, since the dynamical 
time for the assembly to occur, $\lta 1$ Gyr, is 
very much less than the $\sim 10$ Gyr time required 
to create $M_{Fe} \approx 10^8$ $M_{\odot}$ by 
Type Ia supernovae. 
Stars in galaxies undergoing dynamical friction spend most 
of their total lifetimes in the assembled elliptical galaxy, 
not in the assembly itself. 

In this paper we consider an enrichment process 
in which iron from SNIae 
in the central elliptical galaxy is heated by AGN activity
and conveyed by buoyant transport to $r \lta 100$ kpc 
in the surrounding hot gas where it is currently observed.
We show below that the AGN energetics required to 
centrally heat the gas are reasonable and consistent with
other properties of galaxy groups containing hot gas.
Furthermore, to store most or all of the iron from 
Type Ia supernovae over $\sim 10$ Gyrs, it is essential 
that very little of the iron-rich gas cools.
This is possible since 
the failure of cooling flows to cool is now 
widely accepted. 

While we have used the NGC 5044 group 
in our discussion, the same 
argument could be applied to many other similar galaxy groups
and poor clusters with $kT \sim 1$ keV: 
RGH 80 (Xue, B\"ohringer \& Matsushita 2004), 
NGC 507 (Kim \& Fabbiano 2004), 
MKW 4 (O'Sullivan et al. 2003) and
NGC 1399 (Buote 2000a,b). 
In every case the hot gas surrounding the central 
elliptical has an extended region that can be fit 
better with more than one local temperature 
and is enriched 
to $\langle z_{Fe} \rangle \sim 0.5$ solar 
out to $\gta 50$ kpc by the products of Type Ia supernovae. 
The poor clusters 
M87 (Gastaldello \& Molendi 2002) and Centaurus 
(Sanders \& Fabian 2002) also have extended 
multi-temperature, iron-rich regions surrounding
the central elliptical galaxy.
 
\section{Circulation Flow Equations}

We consider a spherically symmetric
cooling flow that flows in to some small 
radius $r_h$ where it is heated.
Plumes and bubbles of heated gas rise buoyantly upstream 
in the cooling flow until their entropy matches that of the 
inflowing gas, provided there is no energy exchange
between the bubbles and the ambient flow.
The rising gas bubbles displace the inflowing gas and reduce the volume
through which it flows.
Larger bubbles rise faster than smaller ones since they 
experience less drag per unit mass 
(cf. Equation 13 of Mathews et al. 2003).
The volume filling factor $f$ available to the inflowing gas
is difficult to estimate because it depends on the 
unknown bubble size distribution. 
Therefore we use observations of
NGC 5044 and RGH 80 as guides to parameterize $f(r)$ 
with the following expression
\begin{equation}
f(r) = 1 - f_h e^{(r_h - r)/r_{2t}}
\end{equation}
where $f_h = 1 - f(r_h)$ 
and $r_{2t} = 30$ kpc is the maximum radius for which the
X-ray spectrum shows two temperature components
(Buote et al. 2003a).
This form for $f(r)$ is also similar to the self-consistent filling
factors we calculated for steady state 
circulation flows (Mathews et al. 2003).
In reality $f$ may be a function of both radius and time, 
but we demonstrate below that the inflowing solutions are
surprisingly insensitive to our choice for $f(r)$, 
and presumably also its time variation. 
The volume filled with inflowing gas in 
a shell of thickness $dr$ is 
\begin{equation}
dV = 4 \pi r^2 f dr = A dr
\end{equation}
where $A$ is the area available to the cooling inflow at radius $r$.

In this first time-dependent calculation of mass recirculation
in cooling flows, we treat the central galaxy as a 
source of heating and iron enrichment at radius $r_h = 5$ kpc.
This is a reasonable approximation since the scale of the 
central galaxy is much less than 
the observed extent $\sim 50 - 100$ kpc of the region of
SNIa-enriched gas surrounding it.
The equation of continuity for the cooling inflow is 
\begin{equation}
{\partial \rho  \over \partial t}
+{\bf \nabla}\cdot \rho  {\bf u}
= {dp \over dV} \left\{ M_*
\left[ \alpha_* +
\alpha_{sn}  \right] + |{\dot M}_h| \right\}.
\end{equation}
The source term on the right represents the rate that 
mass is recirculated per unit volume by 
rapidly buoyant bubbles arriving from the galactic core.
Here 
\begin{equation}
{\dot M}_h  =  4 \pi r_h^2 \rho_h u_h f(r_h) 
\end{equation}
is the rate that mass arrives at the heating radius $r_h$
and $ M_*(\alpha_* + \alpha_{sn})$ is the 
(generally smaller) rate that new 
mass is created in the central galaxy by stellar mass loss
and Type Ia supernovae (SNIa) respectively. 
The specific stellar mass loss rate is 
\begin{equation}
\alpha_*(t) = 4.7\times 10^{-20} (t / t_n)^{-1.3}~~~{\rm s}^{-1}
\end{equation}
where $t_n = 13.7$ Gyrs is the current age of the universe.
The mass loss rate from SNIa is 
\begin{equation}
\alpha_{sn} = 3.17 \times 10^{-20} SNu(t)
\left({M_{sn} \over M_{\odot}}\right) {1 \over \Upsilon_B}
~~~{\rm s}^{-1}
\end{equation}
where
$\Upsilon_B$ is the stellar mass to B-band light ratio
and $M_{sn} = 1.4$ $M_{\odot}$ is the mass ejected in each SNIa event.

The coefficient $dp/dV$ in the source term for the equation of 
continuity is the normalized probability that mass heated at 
radius $r_h$ is transported to a remote volume $dV$ in 
the flow.
In our study of steady state circulation flows we showed that 
the time for gas bubbles heated at $r_h$ to move out to their 
final radius was $\lta 0.15$ of the flow time back to $r_h$.
The time required for heated bubbles to reach their final radius 
is comparable to the dynamical time in the cluster potential
which is short compared to the cooling (or inflow) time;
$t_{dyn} \ll t_{cool}$ is a defining characteristic of 
so-called cooling flows.

The normalization for the circulation probability is 
\begin{equation}
\int_{r_h}^{r_e} {dp \over dV} 4 \pi r^2 f dr
= \int_{r_h}^{r_e} {dp \over dr} {4 \pi r^2 f \over A} dr
= \int_{r_h}^{r_e} {dp \over dr} dr = 1
\end{equation}
where $r_h$ and $r_e$ are the minimum and maximum radii that 
define the region of circulation.
This suggests a simple probability $p(r)$ that depends only
on the radius, $dp/dr = A (dp/dV)$.
For the mass redistribution we consider parameterized functions 
of the form 
\begin{equation}
{dp / dr} = ({dp / dr})_o  (r/r_p)^n e^{-({r/r_p})^m}
= ({dp / dr})_o  \xi^n e^{-\xi^m}
\end{equation}
where 
\begin{equation}
\xi = r/r_p, 
\end{equation}
$r_p$ is an adjustable scale factor and 
$m$ is either 1 or 2 but $n$ may have any positive value. 
The normalization can be expressed in terms of 
incomplete gamma functions, 
\begin{equation}
\left({dp \over dr}\right)_o = 
\left\{ {r_p \over m} \left[
\gamma\left({n+1 \over m}, \xi_e^m\right) 
- \gamma\left({n+1 \over m}, \xi_h^m\right) \right] \right\}^{-1}
\end{equation}
where $\xi_e = r_e /r_p$ and $\xi_h = r_h/r_p$.
The probability distribution has a maximum at
$r_m = r_p (n/m)^{1/m}$.

The density of iron requires a continuity equation of 
its own.
If $n_H$ is the proton number density in the hot gas, 
the density of hydrogen is $\rho_H = n_H m_p$ where
$m_p$ is the proton mass.
The total density is
\begin{equation}
\rho = n_H m_p {5 \mu \over 4 - 3\mu} \equiv \rho_H \phi
\end{equation}
where $\mu = 0.61$ is the mean molecular weight
for full ionization and 
$\phi = 5 \mu /(4 - 3\mu) = 1.41$ allows for the mass of helium.
Since
\begin{equation}
z = {\rho_z \over \rho_H} = \phi {\rho_z \over \rho},
\end{equation}
the density of element $z$ (i.e., iron) 
is $\rho_z = {\rho z / \phi}$.
The gas dynamical evolution of the iron is most easily 
described by regarding the product $\rho z$ as a single 
variable,
\begin{displaymath}
{\partial (\rho z) \over \partial t}
+{\bf \nabla}\cdot (\rho z) {\bf u}
\end{displaymath}
\begin{equation}
~~~= {dp \over dV} \left\{ M_*
\left[ \alpha_* \langle z_* \rangle +
\alpha_{sn} {y_{Ia,Fe} \over M_{sn}} \phi \right] + |{\dot M}_h| z_h
\right\}.
\end{equation}
Here $\langle z_* \rangle = 0.7 z_{Fe,\odot}$ is the adopted 
mean stellar iron abundance in the central galaxy 
(Rickes et al. 2004) and 
we take $z_{Fe,\odot} = 1.83 \times 10^{-3}$ as the solar abundance.
Each SNIa remnant is assumed to contain $y_{Ia,Fe} = 0.7$ solar masses
of iron. 
The production of iron by Type Ia supernovae is about ten times 
greater than that due to stellar mass loss.

The equation of motion is 
\begin{equation}
{\partial {\bf u} \over \partial t}
+{\bf \nabla}\cdot {\bf u}
= - {1 \over \rho} {dP \over dr} - {\bf g}.
\end{equation}
The gravitational acceleration 
$g = G M /r^2$ in the NGC 5044 group 
can be approximated with
\begin{equation}
g = \min[2.4319 \times 10^{-7} r_{kpc}^{-0.75362},
4.762 \times 10^{-7}] ~~~ {\rm cm~s}^{-2}
\end{equation}
which is an excellent approximation in $0.1 < r_{kpc} \lta 500$ kpc
to the mass profile discussed by Buote, Brighenti \& Mathews (2004), 
i.e. a de Vaucouleurs stellar mass of 
$M_{*t} = 3.4 \times 10^{11}$ $M_{\odot}$ with radius 
$R_e = 10.0$ kpc and a dark NFW halo
(Navarro, Frenk \& White 1997) 
of mass $4 \times 10^{13}$ $M_{\odot}$ with concentration
$c = 13$.

Although the temperature and density in rising 
bubbles should both match those in the surrounding gas 
when their entropies become equal,
the mass added by recirculated gas in some 
remote volume $dV$  
must cause the local gas density to increase with time. 
If the density increases in some region of the flow, 
the increased radiative cooling there increases the 
density further.
Eventually, 
the gas may cool catastrophically by runaway radiative losses.
The dense, cooled gas then falls in the gravitational potential,
and after time becomes
supersonic, increasing $|{\dot M}_h|$ enormously.

To avoid these undesirable consequences, 
it is necessary to heat the inflowing gas 
as it moves past the rising bubbles. 
This heating can be due to $Pdv$ work done by the 
expanding bubbles, the power expended by the drag force 
on the rising bubbles, and thermal mixing of some bubble 
gas with the local inflow.
The thermal evolution of the bubbles may not be purely adiabatic.
In this discussion we wish to avoid a detailed discussion 
of these complicated bubble-flow interactions and simply 
assume that a power
$\epsilon_h L_h$ is transferred to the cooling flow
where $L_h$ is a characteristic luminosity expended in heating 
gas at $r_h$ and $\epsilon_h$ is a dimensionless factor 
of order unity.
The magnitude of this heating luminosity will be 
justified below.
We also assume for simplicity that this distributed heating 
is transferred to the cooling flow with
the same probability distribution
$dp/dV$ as the mass recirculation.
Obviously, some very
complicated bubble physics is subsumed with this assumption, some of
which were discussed in our previous paper. 
However, by combining these detailed interactions into 
a single energy source term, we can explore 
new gas dynamical possibilities that must be verified
by more detailed calculations in the future.
The evolution of the energy density
$e = P/(\gamma - 1)$, including this heating and radiation 
losses, is described by
\begin{displaymath}
{\partial e \over \partial t} + {\bf \nabla \cdot u} e
= - P {\bf \nabla \cdot u} 
-{\rho^2 \over m_p^2} \Lambda 
\end{displaymath}
\begin{equation}
~~~+{ 3 k T \over 2 \mu m_p}
{dp \over dV} \left\{ M_*
\left[ \alpha_* +
\alpha_{sn}  \right] + |{\dot M}_h| \right\}
+ \epsilon_h L_h {dp \over dV}
\end{equation}
where $\epsilon_h$ is of order unity 
and $\Lambda(T,z)$ is the usual coefficient for 
optically thin radiative cooling
(Sutherland \& Dopita 1993).
The terms on the right hand side represent respectively 
compressive heating, cooling by radiative losses, 
addition of thermal energy density associated 
with the (non-heating) mass deposition and 
genuine heating provided by the mass deposition. 
When $\epsilon_h = 0$ the recirculated mass is assumed not 
to heat or cool the gas, 
so the specific thermal energy $3kT/2\mu m_p$ contributed by 
the recirculated gas is identical to that of the local gas.
Consequently, the energy density equation 
must have a source term analogous to that in Equation (5).
This third term on the right side of Equation (18) has
little effect on the solutions.

While the spatial distribution $dp/dV$ of the bubble heating term 
$\epsilon_h L_h (dp/dV)$ need not 
be identical to that of the recirculated mass,
it is the simplest assumption
that involves the fewest number of additional parameters 
and seems appropriate for this initial study of
time-dependent circulation.
The power expended in heating the gas at $r_h$ is 
\begin{equation}
L_h = [M_* (\alpha_* + \alpha_{sn}) + |{\dot M}_h|]
{3 k T_h \over 2 \mu m_p} \langle h \rangle
\end{equation}
where $T_h = T(r_h)$ is the temperature of the inflowing 
gas at $r_h$ and $\langle h \rangle$ is a dimensionless mean 
heating factor required to recirculate the gas.

The mean heating factor $\langle h \rangle$ can be estimated 
by examining how the incoming gas is heated at $r_h$ 
and assuming adiabatic bubble evolution.
Bubbles are formed at $r_h$ with temperature
\begin{equation}
T_{bh} = h T_h
\end{equation}
where $T_h = T(r_h)$ is the cooling flow temperature
at the heating radius.
[Virialized gas lost from stars 
in the central galaxy is also assumed
to have a pre-heating temperature $\sim T_h$; 
in general $|{\dot M}_h|$ is considerably larger 
than $M_* (\alpha_* + \alpha_{sn})$.]
The bubbles are always in pressure balance
with the surrounding gas,
\begin{equation}
T_b = {\rho T \over \rho_b}.
\end{equation}
and so 
\begin{equation}
\rho_{bh} = \rho_h/h.
\end{equation}
As bubbles move out from $r_h$, their entropy is assumed to
be constant,
\begin{equation}
T_b = T_{bh} \left( { \rho_b \over \rho_{bh} }\right)^{\gamma - 1}
= T_h h^{\gamma} \left( {\rho_b \over \rho_h} \right)^{\gamma - 1}.
\end{equation}
Some deviation from constant entropy is expected if the
bubbles radiate or if they lose mechanical energy in heating
the surrounding gas, 
so our adiabatic assumption is a limiting case.
Eliminating $T_b$ from the equations above 
gives an equation for the bubble density at $r > r_h$
\begin{equation}
{\rho_b \over \rho_h} =
{1 \over h} \left( { \rho \over \rho_h}
{T \over T_h } \right)^{1 \over \gamma}
\end{equation}
and, if the entropy is conserved, 
\begin{equation}
{T_b \over T_h} = h
\left( { \rho \over \rho_h} 
{ T \over T_h} \right)^{\gamma - 1 \over \gamma}.
\end{equation}

Adiabatically rising bubbles expand and cool. 
By examination of the last two equations above, 
as outflowing bubbles of constant specific entropy 
approach the radius in the flow where
the flow entropy matches that of the bubbles, 
$S_b \rightarrow S(r)$,
the densities and temperatures also become equal,
i.e. $\rho_b \rightarrow \rho$ and $T_b \rightarrow T$.
The specific thermal energy of gas adiabatically transported 
in this manner to distant regions of the flow 
is identical to that in the surrounding gas and  
no heating occurs in this idealized approximation.
We shall see below, however, that the 
increase in gas mass in the flow due to the deposition of 
outflowing gas actually 
results in a higher gas density, which is further increased 
by (non-adiabatic) radiative cooling. 
Alternatively, if bubbles of hot gas thermally merge 
with their environment at some smaller radius 
where $S_b > S(r)$, the surrounding flow will be heated.

From the last two equations above 
the entropy $S = T/\rho^{\gamma -1}$
of heated bubbles at $r_h$ is
\begin{equation}
S_b = h^{\gamma}S_h
\end{equation}
where $S_h = S(r_h)$.
This last equation associates the heating parameter $h$
with a unique bubble entropy and therefore with a unique
radius $r$ in the cooling flow, assuming that the
entropy of the cooling inflow increases monotonically with radius 
and that the bubbles remain adiabatic.
In NGC 5044 and other similar cooling flows,
as well as for the circulation flows considered here, 
$S \propto r$ is a good approximation, so 
\begin{equation}
h = \left( {S \over S_h} \right)^{3 / 5}
= \left( {r \over r_h} \right)^{3 / 5}
~~~{\rm or} ~~~ r = r_h h^{5 / 3}.
\end{equation}
where we assume $\gamma = 5/3$.
The mean heating factor is approximately 
\begin{displaymath}
\langle h \rangle 
= \int_{r_h}^{r_e} h {dp \over dV} 4 \pi r^2 dr
= {1 \over b(n,m)}
\int_{\xi_h}^{\xi_e} e^{{-\xi}^m}
 \left( {\xi \over \xi_h} \right)^{3 / 5}
\xi^n d \xi
\end{displaymath}
\begin{equation}
= {1 \over b(n,m) \xi_h^{3/5}}
\left[
\gamma\left({n+1.6 \over m}, \xi_e^m\right)
- \gamma\left({n+1.6 \over m}, \xi_h^m\right) 
\right]
\end{equation}
where $b(n,m)$ is the quantity in square brackets in 
Equation (12).
Figure 1 shows a sample of normalized circulation 
probabilities $dp/dr$ that we consider below with each 
curve labeled with the parameters $m,~n,~r_{p,kpc}$ 
and the mean heating factor $\langle h \rangle$. 

In our previous paper on steady state circulation flows 
we considered in detail the dynamics of buoyant bubbles, 
including the exchange of momentum between 
the cooling inflow and the counter-streaming heated bubbles
(Mathews et al. 2003).
We showed that this momentum exchange 
is not particularly important for a wide range of bubble sizes. 
In addition we discussed in detail the contribution of 
the heated gas to the overall X-ray emission. 
The X-ray emission from rising bubbles is small but 
observable, but emission from the hottest gas that 
ends up at the largest radii, is necessarily the lowest 
because of its low density. 
In the standard single-temperature interpretation of the 
X-ray emission from NGC 5044 and other similar flows,
the observed 
temperature is dominated by that of the dense inflowing 
gas, as we have shown in our previous paper on 
circulation flows (Mathews et al. 2003).
Finally, in this earlier paper we discussed in detail the 
various heating processes by which the rising bubbles can 
share energy with the ambient gas. 
We will see below that energy exchange between 
gas flowing in opposite directions is critical to the 
success of time-dependent circulation flows.

\section{Results}

\subsection{Pure Cooling Flows}

Gas dynamical solutions are found by 
solving Equations (5), (15), (16) and (18) 
in spherical symmetry using a 1D ZEUS-like Eulerian code 
(Stone \& Norman 1992).
We employ a computational grid 
of 100 logarithmically increasing zones between 
$r_h = 5$ kpc and $r_e = 500$ kpc 
in which the first zone is 215 pc in width.
Outflow boundary conditions are imposed at $r_h$
and the boundary at $r_e$ is fixed and reflecting since it 
is far beyond the cooling radius ($\sim 60$ kpc)
where the radial flow is negligible.

Our first objective is to establish a 
dynamically relaxed cooling flow 
without mass recirculation, i.e. $dp/dV = 0$.
This flow can serve as a smooth starting point for 
flows with circulation.
Solutions with $dp/dV = 0$ also allow us to explore 
the effect that various filling factors $f(r)$ 
have on the flow.
The initial conditions for these pure cooling 
(and non-circulating) flows 
are the temperature and density profiles observed in 
NGC 5044 (Buote et al. 2003a).
The gas is initially at rest.
These initial profiles are shown with dotted 
lines in Figure 2.
Three cooling flows are 
calculated for $10^{10}$ years 
between $r_h$ and $r_e$ for 
three fixed filling factors having 
$f(r_h) = 0.3$, 0.5 and 1, but all with 
$r_{2t} = 30$ kpc (see Equation 3).

After $10$ Gyrs the computed 
gas density and temperature profiles, shown 
with solid lines in Figure 2, 
remain in reasonably good agreement with the profiles
observed in NGC 5044 today, shown with dotted lines.
While this is reassuring, it has no particular 
significance since there is no reason to expect
that the profiles observed today should persist
over such a long time, particularly since our 
flows include no heating.
Some modest heating does seem to be required 
in $r \lta 100$ kpc 
where the computed temperature profiles 
are $\sim 0.2 - 0.3$ keV below those observed.
Gas beyond about 100 kpc does not change 
appreciably during the 
entire calculation although information moves at the 
sound speed from $r_h$ to $r_e$ in only 0.95 Gyr.
Both observed and computed temperature profiles $T(r)$ 
exhibit maxima well beyond the optical half-light radius, 
$R_e = 10$ kpc, of the central elliptical galaxy. 
This thermal feature, which is 
observed in the hot gas on all 
scales from galaxies to rich clusters, is a characteristic 
signature of cooling by radiative losses.
The overall approximate agreement of our simple cooling 
flows with NGC 5044 indicates that, to a first 
approximation, the X-ray emission from this group 
is dominated by inflowing gas that resembles a 
conventional cooling flow.

The observed 
entropy, pressure and density profiles plotted in Figure 2 
with dotted lines 
have been interpreted assuming 
that the filling factor is unity. 
However, the computed solid line curves in Figure 2 are the actual 
variables in the gas flowing for each assumed $f(r)$.
Therefore, 
to compare the computed flows (solid lines)
in Figure 2 with observations, 
the computed density profiles would need to be adjusted 
downward by $n_e \rightarrow f(r)^{1/2} n_e$ 
and the entropy upward 
by $S(r) \rightarrow f(r)^{-1/3}S(r)$. 
Both these adjustments tend to 
bring the solid curves into better agreement with
the observations at small radii.

The most useful result in Figure 2 is the insensitivity
of the inflowing gas density and temperature profiles 
to large variations in the filling factor $f(r)$.
To the accuracy that we require for our goals here, 
the computed gas profiles are identical for all three $f(r)$.
The similarity of flows with different filling 
factors is particularly advantageous because 
of the difficulty in determining the filling factor
either from first principles or 
from very detailed numerical calculations; 
fortunately $f(r)$ can be found from observations. 
In our study of steady state circulation flows 
we were able to determine the filling factor for the 
return flow in a self-consistent manner only if 
all bubbles were assumed to have the same mass 
(Mathews et al. 2003). 
However, in view of the insensitivity to $f(r)$, 
none of the conclusions that we 
reach below depends strongly on the particular filling 
factor $f(r)$ chosen or its possible variation with 
time, provided it does not deviate radically from 
the three $f(r)$ illustrated in Figure 2 
that bracket the known observations of galaxy groups.
Therefore, in the remaining flows discussed below we adopt 
a fixed filling factor with parameters $f_h = 0.5$ and 
$r_{2t} = 30$ kpc.

\subsection{Non-heated Cooling Flows with Mass Circulation}

As discussed earlier, we consider the 
possibility that most or all of the gas that flows into 
the central galaxy, as well as gas expelled from evolving stars, 
is heated and buoyantly transported, with its iron, 
into a large volume of the surrounding hot gas. 
Heating near $r_h$ is assumed to be continuous and 
isotropic, creating bubbles that flow out in every direction. 
In this initial study of time-dependent circulation flows we 
exploit the defining property of cooling flows that the 
dynamical time is much less than the cooling time.
Indeed, as we showed in Mathews et al. (2003), the dynamical 
time for heated gas to rise buoyantly is much less than the 
radiative cooling time that governs the incoming flow. 

In this section we describe circulation flows in which 
bubbles of gas heated near radius $r_h = 5$ kpc float outward 
without heating the incoming flow.
In view of the variety of complicated ways that 
rising bubbles can heat the inflowing gas, 
it is useful to consider 
flows with no heating as a useful 
intermediate step toward more realistic flows.
These idealized bubbles rise adiabatically to
their final radius and deposit gas into the
cooling flow at a radius where the entropy
and specific thermal energy
of the bubble are identical to those in the ambient flow.

Non-heated circulation flows can be 
studied by allowing recirculation of mass, 
$dp/dV \ne 0$, but not heat energy, 
i.e. $\epsilon_h$ is set to zero in Equation (18) 
so there is no distributed energy source. 
Each flow (and those discussed subsequently) 
begins at cosmic time $t = 2.65$ Gyrs, 
corresponding to a redshift of 2.5, 
with initial conditions taken 
from the fully-relaxed $f = 1$ solution in Figure 2. 
The spatial evolution of the iron abundance 
$z_{Fe}(r,t)$, is also followed.
We find that such non-heated flows cool catastrophically 
and deviate strongly from the observations. 
Such cooling is not desired in our solutions since 
we wish to retain all of the SNIa iron in the hot gas 
as the observations suggest. 

If the mass circulation pattern $dp/dV$ is too concentrated 
near some radius $r_m$, the increased gas density there
leads to rapid radiative losses and eventual cooling.
Such a case is illustrated in Figure 3 
in which the recirculation and heating parameters 
for $dp/dV$ are
($m,n,r_{p,kpc},\epsilon_h$) = (2,18,10,0).
For this $dp/dV$ the 
recirculated mass is deposited in an annulus 
$\sim 10$ kpc wide located at $\sim 30$ kpc.
As it grows in mass, this shell cools and 
settles inward in the gravitational potential. 
At time 3.6 Gyrs, as shown in Figure 3, 
some gas has cooled to $T \lta 10^5$ K and the flow does not 
subsequently recover from this state.

Therefore, the commonly held 
notion that bubbles of heated gas rise adiabatically 
to some radius where their entropy $S_b$ matches 
that of the local cooling flow $S$ cannot be sustained 
because the local entropy of the flow is reduced 
as recirculated mass is introduced. 
Even if the mass in an annular shell increases at constant temperature, 
the gas density increases and local radiative 
losses are stimulated; both processes decrease the local gas entropy. 
This is an important consideration that leads to the failure
of all models described in this section. 
Our discussion in \S 3 relating 
the heating factor $h$ to a unique radius $r$ must 
necessarily be viewed as a semi-quantitative idealization.

The iron abundance in Figure 3 is distributed within a region 
between the current location of the dense shell 
and the somewhat larger
radius $r_m$ at which $dp/dV$ has a maximum.
A similar localized cooling can develop if the mass recirculation 
is more evenly distributed just 
between $r_h$ and some radius $r_{max}$ less than the cooling radius.
If the mass deposition decreases sharply at $r_{max}$, 
the cooling inflow from $r > r_{max}$ can be locally decelerated 
and compressed near $r_{max}$.
This sort of flow, not illustrated here, can result in 
radiative cooling to low temperatures in a gaseous shell.
Such compressive cooling was anticipated in our 
earlier work on steady circulation flows and 
resembles the cooling galactic drips discussed 
by Mathews (1997).

Circulation flows without distributed heating 
($\epsilon_h = 0$) also fail 
even if the mass redistribution $dp/dV$ is very broad.
Such a flow with parameters 
($m,n,r_{p,kpc},\epsilon_h$) = (1,1.5,30,0)
is illustrated in Figure 4 at time 8.0 Gyrs when the flow 
has cooled catastrophically near $r_h = 5$ kpc. 
In this flow the entropy 
$S = kT/ n_e^{2/3}$ out to 
$\sim160$ kpc has been dramatically reduced because 
of radiative losses associated with the increased density.
As the density increases,
the gas pressure gradient must become more negative 
to balance the gravitational potential, 
causing the gas pressure at small radii to increase.
The gas pressure actually develops a 
maximum just beyond $r_h$ within which gas is forced inward.
The entire gas flow within $\sim160$ kpc is  
approaching that of freefall and the mass flux at 
$r_h$, $|{\dot M}_h|$, has become very large. 
The high inflow velocity and the rapid recirculation of 
gas from $r_h$ to larger radii cause the iron abundance 
to be low and flat-topped as seen in Figure 4.

\subsection{Circulation Flows with Heating}

Fortunately, catastrophic cooling 
is easily eliminated by introducing 
a distributed heating term in the energy equation, 
$\epsilon_h > 0$.
For simplicity, we consider heating 
that is distributed over the flow with the same 
normalized probability distribution $dp/dV$ that describes
the mass circulation. 
Even with this constraint
many circulation flow models are 
in satisfactory agreement with observation.

The time-dependent flows we describe here are often
in a quasi-steady state, stable over long periods of time,
so it is interesting to consider how the interbubble
gas is heated by counter-streaming bubbles in the
steady state limit.
As a matter of principle,
an ensemble of outflowing and expanding bubbles 
does not heat the surrounding gas by performing $PdV$ work
provided conditions are approximately steady state
(Mathews et al. 2003).
Certainly a single bubble that suddenly appears
and rises into an atmosphere
of hot gas displaces its volume in the ambient gas
at every radius
and moves some ambient gas higher in the gravitational potential.
In this case $PdV$ work is done as the cloud expands and
moves outward.
However, a single buoyant cloud is not a steady state
phenomenon, it is inherently time-dependent.
An approximate steady state is possible only if many clouds
are flowing through each shell of volume $dV = 4 \pi r^2 dr$
at any time.
In this case the volume filled with
rising bubbles $(1 - f)dV$ is nearly constant; as bubbles
exit from the surface at $r + dr$, new ones appear at $r$,
approximately conserving the number and total volume of bubbles.
Individual bubbles expand adiabatically as they move
from $r$ to $r+dr$ and their internal energy decreases
accordingly.
However, this expansion does not continuously heat
the interbubble gas since, in the strict steady state sense,
this work was already done by similar bubbles at an infinite
time in the past.
Since the volume of the interbubble gas does not change in
steady flow, $f dV/dt = 0$ and 
the surrounding gas experiences no continuous $PdV/dt$ heating. 
Instead, as bubbles move through $dV$ 
the ambient cooling flow gas is subsonically rearranged
transversely along equipotential surfaces, conserving
its total volume. 
However, some work is done against viscous forces during this 
rearrangement.
As a result, each bubble experiences a drag force as it
moves upstream through the interbubble gas.
Work done against viscous drag
is a continuous source of spatially
distributed heating in the surrounding gas.

Figure 5 illustrates two heated circulation 
flows that evolved without catastrophic radiative  
cooling from 2.65 Gyrs to 13.7 Gyrs and which are in 
reasonable agreement with the observed properties of 
NGC 5044 and other similar X-ray luminous groups. 
The final temperature and iron abundance profiles 
of a flow described with circulation parameters
($m,n,r_{p,kpc},\epsilon_h$) = (1,1.5,45,1.6)
are shown as light solid and dashed lines in Figure 5.
The parameters for this flow were chosen to fit the 
observed temperature profile of NGC 5044, but  
the iron abundance does not exceed about 0.7 
solar, somewhat lower than that observed in this 
group (Buote et al. 2003b).
The iron abundance observed in NGC 5044, which rises 
to $\gta 1$ solar at $r \lta 20$ kc, 
is better fit with the circulation flow 
($m,n,r_{p,kpc},\epsilon_h$) = (1,1.1,41,1.9)
which is shown with heavy solid and dashed 
lines in Figure 5.
Because of the slightly larger heating $\epsilon_h = 1.9$ 
in this flow, the inflow velocities are 
lower as the gas approaches $r_h$ and 
the iron abundance there is correspondingly larger.
For both flows in Figure 5 the total mass of iron
within 100 kpc is about $10^8$ $M_{\odot}$ since we 
use the same SNIa rate as in Equations (1) and (2). 

Evidently, if we relaxed our constraint that the
recirculation distributions $dp/dV$ 
of mass and heat energy are 
identical and allowed them to vary with time, 
it would be possible to match both the   
abundance and gas temperature profiles 
for any observed group. 
The density, pressure, flow velocity and 
entropy profiles for the two flows shown in 
Figure 5 are acceptable fits to the NGC 5044 observations.
Indeed, we find similar flows 
with a range of parameters 
($m,n,r_{p,kpc},\epsilon_h$), 
indicating that our results are robust.

It is particularly significant that the 
computed temperature
profiles $T(r)$ in Figure 5 have maxima and positive 
gradients $dT/dr > 0$ at small $r$. 
In our previous paper on circulation flows
(Mathews et al. 2003) we showed that
the X-ray emission from hotter buoyant clouds of 
much lower density 
does not significantly alter the single phase temperature 
profile established by the denser inflowing gas.
However, in the diffusely heated 1D and 2D cooling flows 
that we previously solved with 
numerical hydrocodes, we invariably found 
monotonically decreasing temperatures $dT/dr < 0$
in the centrally heated region 
(e.g. Brighenti \& Mathews 2002). 
In those calculations we heated all the gas 
throughout a volume near the center of the flow, 
including gas at all densities, 
and our specific objective was not to 
create or follow bubbles in the heated region.
Some larger buoyant bubbles flowed out from 
the heated region but they were 
not resolved in part because of spurious numerical diffusion.
We avoid numerical diffusion here by 
assigning a fraction $1 - f(r)$ of the volume 
to the buoyantly outflowing heated gas.
But the outwardly increasing temperature near the 
center in Figure 5 is made possible 
only by the combined influence of 
the redistribution of heated mass and the cooling inflow. 

The profiles shown in Figure 5 are not momentary transients 
that happen to be similar to the observations at 
time $t_n = 13.7$ Gyrs.
While these flows are fully time-dependent,
the gas density and temperature profiles 
remain highly stable during the last $\sim 6$ Gyrs of 
the calculations.
These flows are therefore quasi-steady and 
the main secular change is the slowly increasing 
iron abundance $z_{Fe}(r)$.
Finally, we emphasize that 
the circulation of mass that we describe here 
cannot be viewed as normal convection, in which 
the entropy profile is flat,
the temperature fluctuations are small, 
and the temperature gradient is negative.

Figure 6 shows the time variation of several global 
parameters for the two circulation flows illustrated 
in Figure 5.
The mass flow rate $|{\dot M}(r_h)|$ at $r_h$,
the stellar mass loss rate
${\dot M}_*$ and the approximate power expended in heating 
the gas at $r_h$, $L_h$, in units of $10^{42}$ erg s$^{-1}$
all slowly decrease with time.
The current heating luminosity, 
$\epsilon_h L_h = 3.5 \times 10^{42}$ erg s$^{-1}$, 
is comparable to the X-ray luminosity 
of NGC 5044, $L_x = 5.5 \times 10^{42}$ erg s$^{-1}$,
and this is expected in order to 
maintain the flow without
radiative cooling to low temperatures.
This level of heating $\epsilon_h L_h$
corresponds to a rather low level 
AGN, which is consistent with other observations 
of NGC 5044. 

In Mathews et al. (2003) we showed that the heating 
of the cooling inflow by bubbles due to 
work done against drag forces and $Pdv$ expansion 
(if applicable) are comparable. 
We now estimate the collective heating by bubbles in 
NGC 5044 due to drag interactions. 
Assuming that the bubble velocity $u_b$ is much greater 
than that of the cooling inflow, 
the terminal velocity is 
$u_b \approx (4 g r_b \alpha/3 \delta)^{1/2}$ where
$\delta \sim 0.5$ is the dimensionless drag coefficient
and $\alpha = 1 - (\rho_b/\rho)$. 
The rate that energy is lost from a single bubble 
of radius $r_b$ is then 
$\ell_d = \delta \rho u_b^3 \pi r_b^2$.
The space density of bubbles is
$n_b \approx 3(1 - f)/4 \pi r_b^3$.
The heating rate due to all bubbles is
$L_d \approx \int 3 (1 - f) \delta (\rho/r_b) u_b^3 
\pi r^2 dr \approx 9 \times 10^{42}$ erg s$^{-1}$,
assuming $\langle h \rangle = 5$, coherent 
bubbles of radius 0.5 kpc formed at $r_h = 5$ kpc 
and taking $\rho(r)$ as the density profile observed in NGC 5044. 
The bubbles are assumed to be in pressure equilibrium 
and the integration extends from $r_h$ to 
the radius of equal entropy where $S_b = S$.
In spite of the obviously approximate nature of this estimate,
it is within a factor of 2 or 3 of the heating luminosities 
$\epsilon_h L_h$ 
required for the flows illustrated in Figure 5. 
Therefore the amount of distributed heating we require 
for successful flows is comparable to that expected from 
bubble-flow interactions.

In our models the heating is continuous, but it
may be possible to create similar circulation flows
with intermittent heating near the center.
It is likely, however, that the intermittency if it 
exists has a rather short period or duty cycle. 
The dynamical outflow time for bubbles of size 
$r_b$ at radius $r$ is $t_b \approx r/u_b$.
For bubbles of radius $r_b = r/5$, at $r = 10$ kpc, 
$u_b \approx 260$ km s$^{-1}$ and the bubble rise time
$t_b \approx 3 \times 10^7$ years is rather short.
If gas is heated periodically, the period 
would have to be less than this  
and may in fact be very much less.
Hot bubbles appear in the Chandra image of NGC 5044
well within 10 kpc from the
center, and the X-ray image of NGC 5044 is similar to
those of other elliptical galaxies with comparable $L_B$. 
If mass circulation is a common feature of all 
flows of this type, the implication is that there
are no truly quiescent supermassive black holes.

Finally, we note that the heating due to Type Ia supernovae 
is small compared to $L_h$. 
For example, at time $t = t_n$ the total SNIa 
heating luminosity 
is $L_{Ia} \approx (M_{*t}/M_{sn})E_{sn} \alpha_{sn}$ 
ergs s$^{-1}$ where $E_{sn} = 10^{51}$ ergs is released 
with each supernova.
Evaluating at $t_n$, we find 
$L_{Ia} \approx 2.3 \times 10^{41}$ erg s$^{-1}$ which 
is very much less than $\epsilon_h L_h$.
For simplicity we have ignored supernova heating 
in Equation (18).

\subsection{Central Iron Abundance Minimum}

One of the strangest features 
occasionally observed in cooling flows is a 
central minimum in the iron abundance
(Johnstone, et al 2002;
Sanders \& Fabian 2002;
Schmidt, Fabian \& Sanders 2002;
Blanton, Sarazin \& McNamara 2003;
Dupke \& White 2003).
When present, these minima typically lie within 
the central $\sim 10 - 50$ kpc of the flows, i.e. 
where conditions are dominated by the central elliptical 
galaxy.
Such minima -- or a central flattening in $z_{Fe}(r)$ 
which is even more common -- would not be expected 
in conventional cooling flows since the local frequency 
of iron-producing Type Ia supernovae should increase 
with the stellar density toward the very center of the flow.
(The stellar density is more centrally peaked 
than the gas density.)
Efforts to understand the central dips in the iron 
abundance in terms of differential cooling of 
abundance inhomogeneities 
have been discussed by Morris \& Fabian (2003).

In the circulation flows described here it is possible 
to produce small central iron abundance minima if the 
redistribution probability $dp/dV$ varies with time.
For example, a small central dip in the iron abundance 
results if the iron (and mass) deposition $dp/dV$ is concentrated 
near $40-50$ kpc, just preceding the flow time 
for gas in this region to return to $r_h$ by 
the current time $t = 13.7$ Gyrs.
Central regions with very low iron abundance are 
difficult to form in circulation flows 
because (1) the cooling inflow 
necessary to explain the observed gas temperature profile 
continuously recirculates iron-enriched gas within $\sim50$ kpc
and, since most of the SNIa iron is produced at early times, 
(2) relatively little iron is available during the 
last cooling inflow time  
to create a strong reversal in the iron abundance profile. 
For these same reasons, central iron deficiencies are not 
easy to produce by allowing the iron mass and energy 
redistribution probabilities $dp/dV$ to differ.
Nevertheless, because of the continuous radial recirculation 
it is quite natural for the central  
gas iron abundance in circulation flows to become rather constant 
within $\sim30$ kpc 
where the SNIa-producing stars are very strongly peaked.
Similar flat or slowly varying iron abundance cores 
are commonly observed in many groups and clusters.

\section{Additional Remarks}

Although the heated flows we describe here are generally 
successful, much additional work will be necessary before
they can be fully accepted.
We have not identified the physical mechanism(s) 
that heat the gas near the center, but they are likely to be 
associated with the central black hole.
These heating processes may be
related to the common difficulty in observing hot gas in close
proximity to the central black hole, 
constraining the accretion rate to less than 10 percent 
of the Bondi rate (Loewenstein et al. 2001).
If gas is heated near the black hole, 
less accretion is expected.
It is likely that cosmic rays and magnetic fields 
are relevant since the cores of almost all massive elliptical 
galaxies are sites of low luminosity nonthermal radio emission
(e. g. Sadler et al. 1994).
Powerful radio jets are typically found in rich clusters, 
not in (more numerous) small groups that contain 
central ellipticals of comparable optical luminosity.
Jets that deposit energy asymmetrically 
and to very large radii are an unlikely heating source for 
groups like NGC 5044 and RGH 80.
By contrast, the heating required in our circulation flows 
is isotropic and requires a modest, approximately continuous 
AGN power.

In our discussion of steady state circulation flows  
(Mathews et al. 2003), we describe limits on the size and mass 
of bubbles if heating occurs at some radius $r_h$ in the flow.
Provided they remain coherent, larger bubbles 
can move to larger radii, but the size of new bubbles 
cannot exceed the radius $r_h$ where they are formed 
and this in turn limits the global rate that mass can be 
buoyantly transported outward. 
We discussed these difficulties in the context 
of steady state circulation and they did 
not critically impair the circulation model. 
In this first paper on time-dependent circulation 
we have not discussed again all these important considerations. 
The insensitivity of circulation flows to the filling factor
also lessens the urgency for a fully self-consistent model
for the rising bubbles, even if that were possible.
Nevertheless, as we computed these circulation flows,
our concept of the heating radius $r_h$ became more flexible. 
We consider $r_h$ to be  
only a representative radius where the gas is heated.
Larger (smaller) clouds may be heated and formed at larger 
(smaller) radii.
In reality the heating probably occurs throughout a volume that 
may vary with time. 

A critical measure of the success of circulation flows 
is the apparent rise in the gas temperature toward a 
maximum well outside the central elliptical galaxy. 
This $T(r)$ is possible only if the emission from 
the high density cooling inflow dominates the observed 
temperature profile in the inner flow, even when the mass flux 
carried by the rising bubbles transports 
nearly the same mass outward.
We showed that this is indeed the case for steady state 
circulation flows (Mathews et al. 2003).
To see why this is also true for time-dependent circulations, 
we estimate the ratio of the bubble to cooling flow 
bolometric X-ray 
emissivities at $r = r_h$ at time $t = t_n$, 
\begin{equation}
{\epsilon_{bh} \over \epsilon_{h}} =
\left( {\rho_{bh} \over \rho_h} \right)^2
{\Lambda(T_{bh}) \over \Lambda(T_h)}
{1 - f(r_h) \over f(r_h)}.
\end{equation}
The bubble density at $r_h$ can be estimated from the 
steady state equation for mass conservation,
\begin{equation}
\rho_{bh} \approx { (|{\dot M}_h| + {\dot M}_*) \over 
4 \pi r_h^2 [1 - f(r_h)]} {1 \over u_{bh}}
\end{equation}
where $u_{bh}$ is the typical velocity of rising bubbles 
near $r_h$.
Using the flow parameters for the 
($m,n,r_{p,kpc},\epsilon_h$) = (1,1.5,30,2.0)
flow at time $t_n$, we find
$\rho_{bh} = 2.3 \times 10^{-26}/u_{bh2}$ gm cm$^{-3}$ 
where $u_{bh2}$ is the bubble velocity in units of 
100 km s$^{-1}$. 
The bubble temperature 
$T_{bh} = 1.5 \times 10^7 u_{bh2}$ K follows from 
pressure equilibrium. 
With these values we find
$\epsilon_{bh}/\epsilon_h \approx 0.06/u_{bh2}^2 
\approx 0.007 \ll 1$ if $u_{bh2} = 3$, a typical 
initial bubble velocity in the steady flows 
of Mathews et al. (2003). 
Therefore, emission from the cooling inflow 
dominates the observed $T(r)$ 
and the bolometric X-ray emission
near the center of this flow.
X-ray spectral features from heated gas  
may be emitted mostly by somewhat denser gas at intermediate 
temperatures, $T_h < T < T_{bh}$.

There are two stages of heating in our circulation flows.
Near the center $r \sim r_h$ we assume a strong and 
concentrated heating by an AGN that rather sharply raises the 
entropy, creating buoyant bubbles of hot gas. 
This is regarded as the primary heating mechanism.
A second heating process occurs when 
the rising bubbles share 
some of this thermal energy with the inflowing gas 
over a large region of the flow.
Without this distributed heating the entropy profile 
in the gas fails to match the observations (as in Figure 4). 
Distributed heating occurs because of bubble-flow drag or 
because some of the outflowing gas thermally merges
with the incoming flow before it reaches the
radius where the initial entropy of the bubble
equals that of the ambient flow.
This distributed heating helps determine the location of the 
temperature maximum observed in the flow. 
Finally, it is not surprising that the central heating exceeds
$L_h$ by some factor $\epsilon_h \sim 2.0$ because
(1) the mean heating factor $\langle h \rangle$ is underestimated
if some of the bubble energy is prematurely transferred to 
the inflowing gas, (2) some of the bubble energy is 
lost by radiation and this is necessary to explain 
the superiority of two-temperature fits to the X-ray spectra, 
and (3) the kinetic energy acquired by the bubbles 
is extracted from the galaxy-group potential 
and is independent of $L_h$. 

X-ray observations of the Perseus cluster show buoyant 
cavities that have moved through distances that 
are 10 - 15 times larger than that of the cavity diameters
(e.g. Fabian et al. 2000).
The coherence of these large bubbles is remarkable, and 
they appear to be transporting mass and heated gas to 
large radii in the flow. 
Since small buoyant regions experience more drag as 
they move upstream through the cooling flow, 
their rate of progress may not be sufficient to 
convey the mass and energy that we require in the 
models described here. 
Clearly, observational and computational 
studies of the physical coherence of heated 
regions will be necessary to fully validate circulation flows. 
The dynamical coherence of heated bubbles may 
result from internal magnetic fields that are force free.

It is possible that some gas cools.
For NGC 5044, $r_h = 5$ kpc, taken literally, 
is about $R_e/2$ so we expect 
$\sim 0.1$ $M_{\odot}$ yr$^{-1}$ from stellar mass loss 
within $r_h$. 
Even if this gas cools radiatively in the usual manner, 
such a small cooling rate would not 
have been detected in our XMM observations
(Buote et al. 2003a).
However, cooling in this core region is likely to 
be accelerated by electron-dust grain cooling and 
emit X-rays at a much reduced level
(Mathews \& Brighenti 2003).
Central (dust-enhanced) cooling may consume $\sim10$ percent 
of the iron produced by SNIae.

\section{Conclusions}

In this paper we describe a new model for the dynamical 
evolution of hot gas in groups and clusters 
of galaxies in which gas flows in both radial 
directions simultaneously.
The incoming gas resembles a traditional cooling flow
and dominates the X-ray emission.
The outgoing gas consists of an ensemble of heated 
buoyant bubbles that rise to distant regions 
where they merge with the inflowing gas. 
We have not calculated the dynamics of the rising bubbles 
of heated gas in detail, and this must be an objective 
for the future.
Instead, we constructed a schematic time-dependent model 
guided in part by observations and by the detailed 
bubble-flow interactions described in 
our previous study of steady state circulation flows 
(Mathews et al. 2003).
It is noteworthy and fortunate 
that the density and temperature profiles 
of the cooling, inflowing gas 
are not sensitive to the filling factor $f(r)$, 
which is easier to determine reliably from the observations 
than from dynamical models for the bubbles. 
This insensitivity may explain the approximate
similarity of $T(r)$ and $n_e(r)$ profiles 
in clusters of all sizes.

We have shown here that the steady state circulation flows 
we proposed earlier (Mathews et al. 2003) 
can be generalized to include time variation.
An essential feature of successful circulation flows is that 
both mass and energy must be distributed from the center 
throughout a large volume of the flow within the cooling radius. 
The outward increase in gas temperature 
near the center of the flow,
a characteristic feature of most galaxy clusters, 
results largely from 
a conventional cooling inflow, modified somewhat by additional energy 
received from locally outflowing bubbles.
After evolving for many Gyrs, 
circulation flows are in accord with the important 
observation that most or all of the iron produced 
by Type Ia supernovae in the central elliptical galaxy
has been stored in a region typically 
extending to $\sim 100$ kpc.
From the perspective of the circulation flow hypothesis,  
a record of the enrichment-heating history of groups 
is retained in currently observed iron and temperature profiles of 
the hot gas and it may be possible to retrieve this 
information by comparison with more accurate circulation flows. 

The central peak in the radial iron distribution 
within $\sim 50$ kpc is very 
sensitive to the (negative) cooling inflow velocity at small 
radii. 
If this velocity is high, the iron abundance remains 
low and flat-topped. 
Iron abundance profiles that peak sharply toward the center,
such as that 
observed in NGC 5044, can occur only if the flow
velocity is reduced to a very low level by 
(recent) large heating near the center.
We have shown that 
the central iron peak is more pronounced 
when the mass redistribution and heating are more strongly
peaked at smaller radii.

Concentrated peaks in the iron abundance within $\sim 100$ 
kpc are also typically observed in many rich clusters:
Sersic 159-03 (Kaastra et al 2001),
Cygnus A (Smith et al. 2002), 
Abell 1795 (Ettori et al. 2002) and 
Abell 2029 (Lewis, Stocke \& Buote 2002). 
We speculate that these regions are also mostly enriched 
by the circulation of iron from Type Ia 
supernovae in the central galaxies 
and there is some observational support for this  
(e.g. Ettori et al. 2002). 
If so,
this may indicate a characteristic heating luminosity 
$L_h \sim 10^{42} - 10^{43}$ ergs s$^{-1}$ for all 
clusters. 
However, 
if there is an upper limit on the heating luminosity $L_h$, 
perhaps set by the physical conditions near the 
central black hole, 
it is possible that the mass inflow rate 
$|{\dot M}|$ in rich clusters is too large 
for the inflowing gas 
to be heated in the way we describe here. 
This may give rise to an additional more
violent energy releases in the form of powerful radio jets 
which are rare in galaxy groups that have more modest 
mass cooling rates $|{\dot M}|$.

We believe the circulation flows described
here are the first gasdynamic, long-term evolutionary 
models that are in good agreement 
with all essential features of the hot gas: 
the gas temperature maximum at several $R_e$, 
the approximately linear rise of entropy with radius, 
the dominance of SNIa products in the X-ray spectra, 
the total iron mass within 100 kpc and its radial distribution.
The relative constancy of these observed profiles
during many Gyrs of our calculated flows  
is consistent with the similarity of the observed 
X-ray properties among galaxy groups and clusters.
Finally, little or no gas cools in circulation models, 
and this is consistent with the much decreased or absence 
of cooling gas in X-ray spectra of 
both galaxy groups and clusters.

\vskip.4in
Studies of the evolution of hot gas in elliptical galaxies
at UC Santa Cruz are supported by
NASA grants NAG 5-8409 \& ATP02-0122-0079 and NSF grants  
AST-9802994 \& AST-0098351 for which we are very grateful.



\clearpage
\begin{figure}
\includegraphics[bb=90 216 522 569,angle=270]
{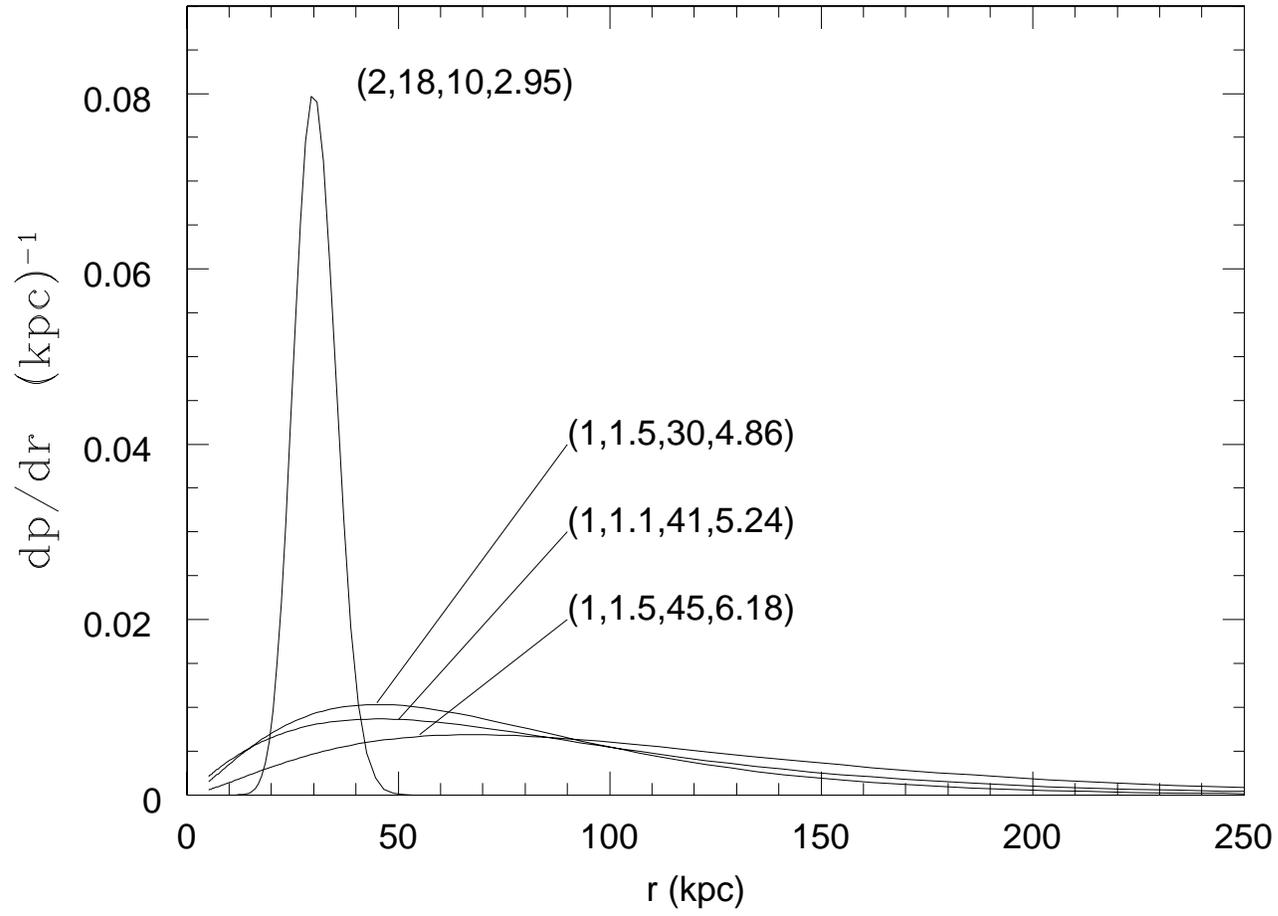}
\vskip.7in
\caption{
Normalized mass redistribution probability functions $dp/dr$ 
each labeled with ($m,n,r_{p,kpc},\langle h \rangle$).
}
\label{fig1}
\end{figure}

\clearpage
\begin{figure}
\includegraphics[bb=90 216 522 569,angle=270]
{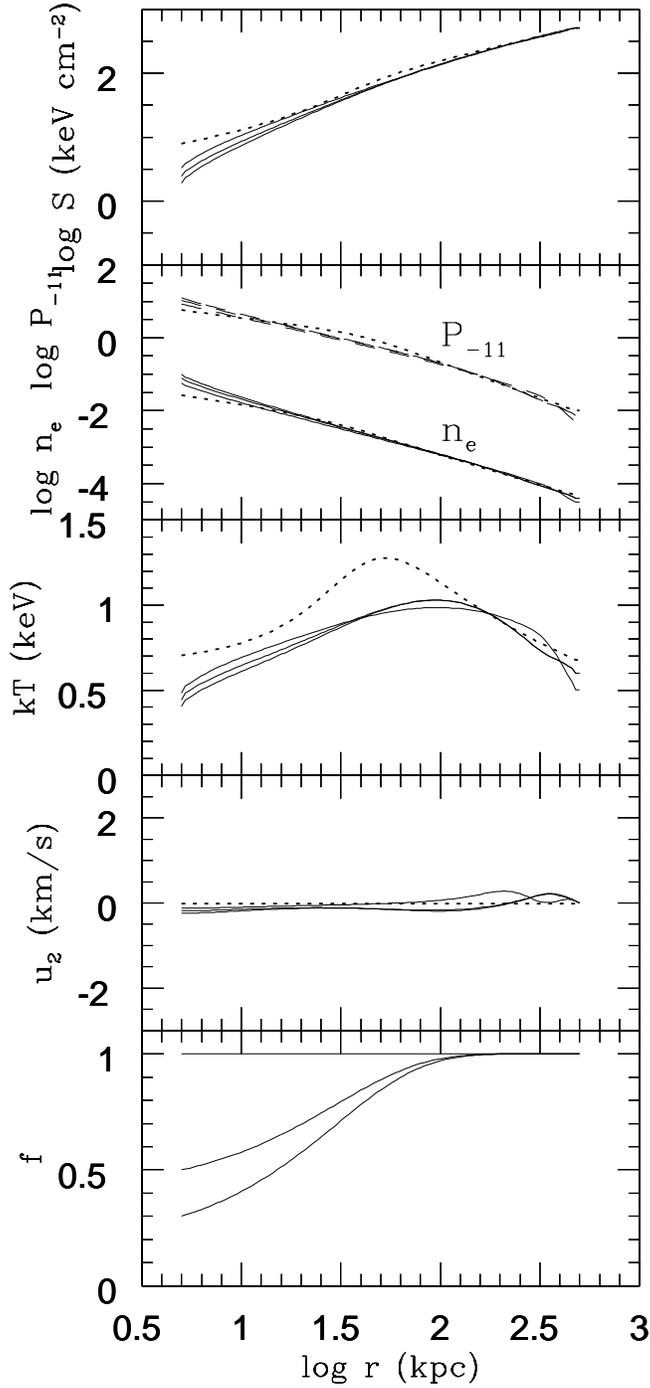}
\vskip.7in
\caption{
Pure cooling flows (solid lines) after evolving for 10 Gyrs
with three different filling factor profiles $f(r)$.
The dotted lines show the
density and temperature profiles observed in NGC 5044
that serve as initial
conditions for the computed profiles; 
the dotted line pressure and entropy profiles are derived 
from the observations.
The filling factors correspond to
$f(r_h) = 0.3$, 0.5 and 1.0, all with $r_{2t} = 30$ kpc.
In descending order from the top
the panels show profiles for
the gas entropy $S = kT/n_e^{2/3}$,
the gas density $n_e$ (solid lines) and pressure $P$
(dashed lines) in units of $10^{-11}$ dyne,
the gas temperature $kT$,
the gas velocity $u_2 = u/(100~{\rm km~ s}^{-1})$.
and the flow filling factor $f$.
The low amplitude velocity profiles
with $f(r_h) = 0.3$ and $f_h = 0.5$ are fortuitously identical.
}
\label{fig2}
\end{figure}

\clearpage
\begin{figure}
\includegraphics[bb=90 216 522 569,angle=270]
{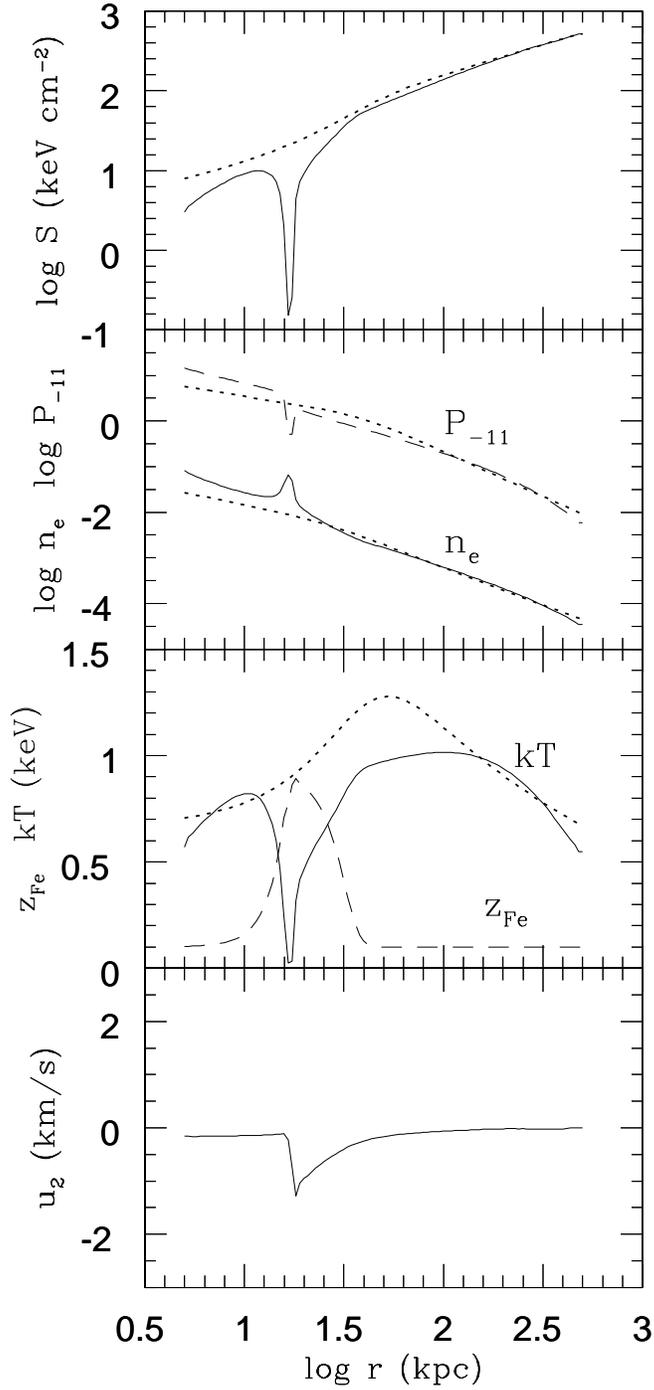}
\vskip.7in
\caption{
The solid lines show a non-heated circulation flow
in which the mass recirculation probability $dp/dV$ is
concentrated near $r_m = 30$ kpc, 
($m,n,r_{p,kpc},\epsilon_h$) = (2,18,10,0).
The flow began at cosmic time $t = 2.65$ Gyrs
and is shown at $t = 3.6$ Gyrs when severe cooling occurs.
The dashed lines show the gas pressure and the
iron abundance in solar units.
The dotted lines show the observed $n_e(r)$ and
$T(r)$ profiles for NGC 5044.
}
\label{fig3}
\end{figure}

\clearpage
\begin{figure}
\includegraphics[bb=90 216 522 569,angle=270]
{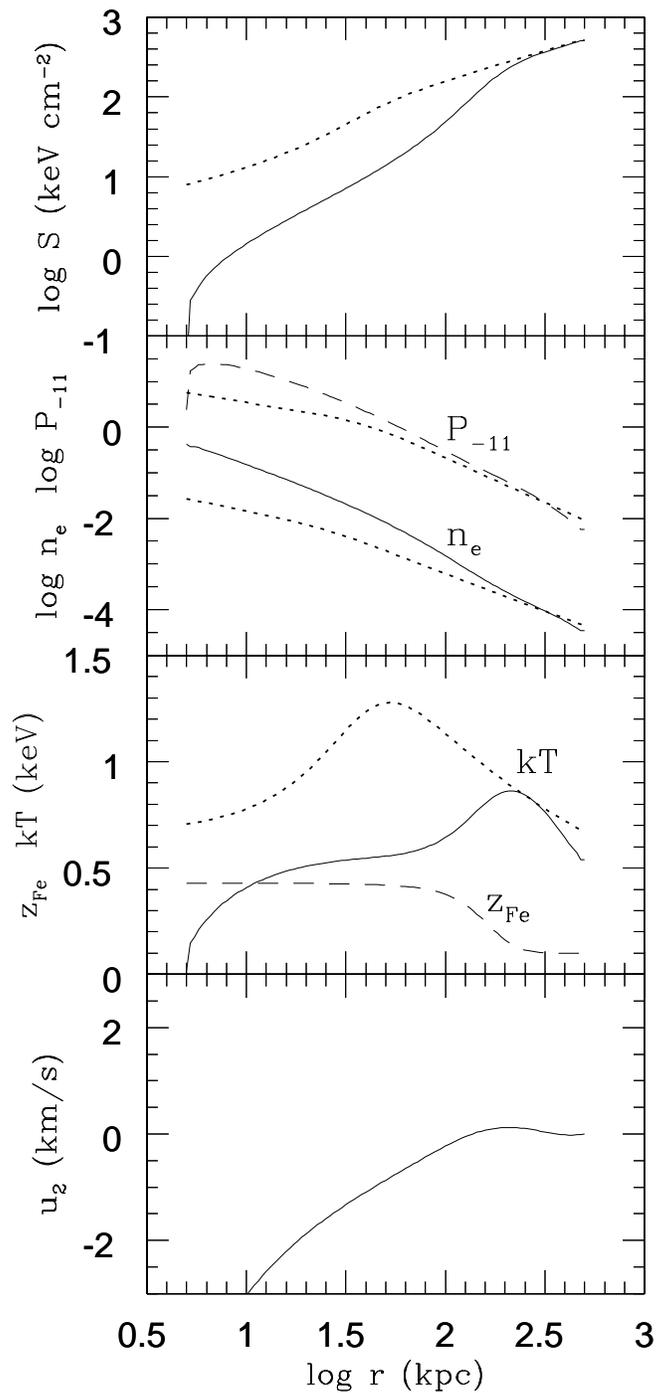}
\vskip.7in
\caption{
The solid lines show a non-heated circulation flow
in which the mass recirculation probability $dp/dV$
is a broad function extending out to several hundred kpc,
($m,n,r_{p,kpc},\epsilon_h$) = (1,1.5,30,0).
The flow is shown at time $t = 8.0$ Gyrs when catastrophic
cooling occurred at the inner radius.
The dashed lines show the gas pressure and the
iron abundance in solar units.
The dotted lines show the observed $n_e(r)$ and
$T(r)$ profiles for NGC 5044.
}
\label{fig4}
\end{figure}

\clearpage
\begin{figure}
\includegraphics[bb=90 216 522 569,angle=270]
{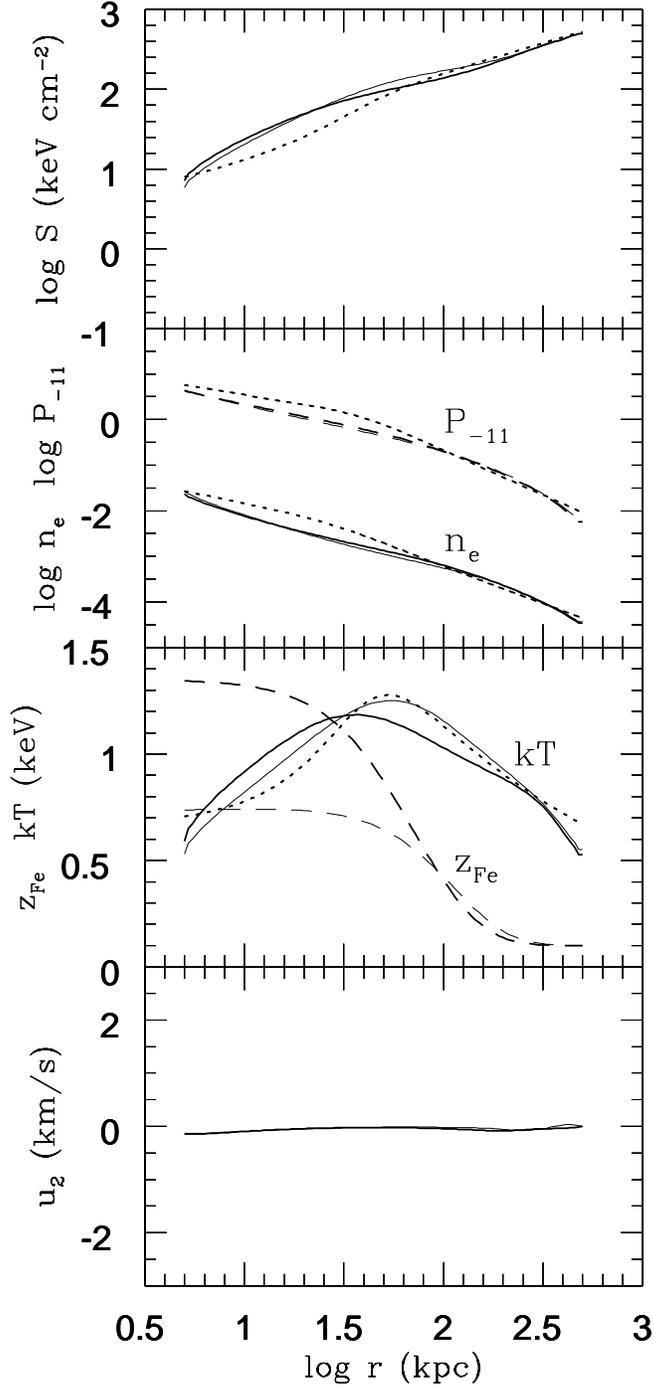}
\vskip.7in
\caption{
The solid lines show two heated circulation flows
with slightly different mass recirculation probabilities $dp/dV$
that give good agreement at time $t = 13.7$ Gyrs with
the observed iron abundance and temperature profiles
observed in NGC 5044.
The dashed lines show the gas pressure and the
iron abundance in solar units.
Heavy and light lines correspond to probabilities
$dp/dV$ defined by
($m,n,r_{p,kpc},\epsilon_h$) = (1,1.1,41,1.9) 
and
($m,n,r_{p,kpc},\epsilon_h$) = (1,1.5,45,1.6) 
respectively.
The dotted lines show the observed $n_e(r)$ and
$T(r)$ profiles for NGC 5044.
}
\label{fig5}
\end{figure}

\clearpage
\begin{figure}
\includegraphics[bb=90 216 522 569,angle=270]
{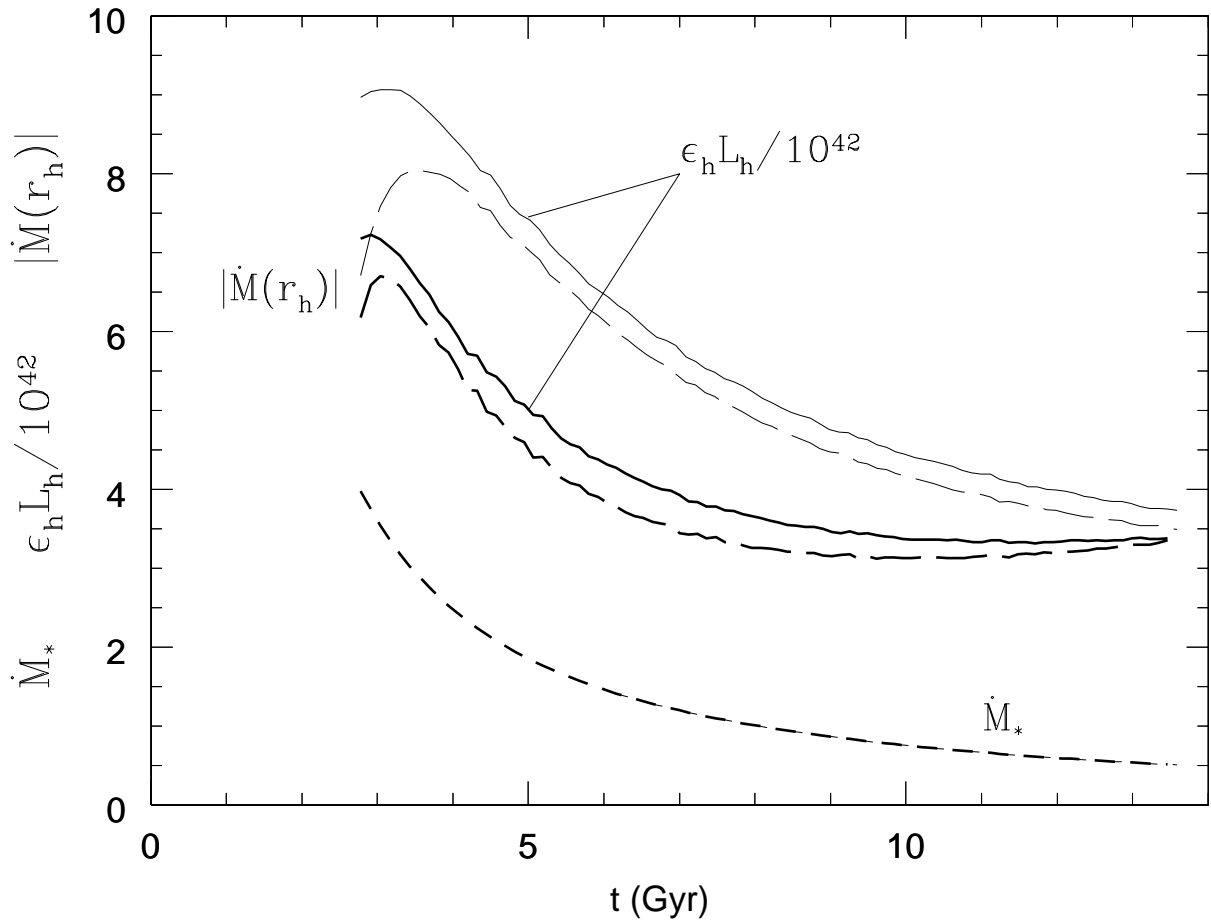}
\vskip.7in
\caption{
Variation with time of
several properties of heated flows with parameters 
($m,n,r_{p,kpc},\epsilon_h$) = (1,1.1,41,1.9) (light lines)
and (1,1.5,45,1.6) (heavy lines).
Shown are the mass inflow rate at $r_h$
$|{\dot M}(r_h)|$ in $M_{\odot}$ yr$^{-1}$
(long dashed lines),
the total stellar mass loss rate ${\dot M}_*$ in
$M_{\odot}$ yr$^{-1}$ in the central E galaxy
(short dashed lines),
and the heating luminosity $\epsilon_hL_h$ 
in units of $10^{42}$ ergs s$^{-1}$ (solid lines).
}
\label{fig6}
\end{figure}

\end{document}